\documentclass[12pt]{article}
\usepackage{amssymb}
\usepackage{amsmath}
\usepackage{contour}
\usepackage{graphicx}
\contourlength{0.3pt}
\contournumber{32}

\newcommand{\hollow}[1]{%
	\text{\textcolor{white}{\contour{black}{#1}}}}

\renewcommand{\theequation}{\arabic{section}.\arabic{equation}}

\begin{document}



\def\a{\alpha}
\def\b{\beta}
\def\d{\delta}
\def\e{\epsilon}
\def\g{\gamma}
\def\h{\mathfrak{h}}
\def\k{\kappa}
\def\l{\lambda}
\def\o{\omega}
\def\p{\wp}
\def\r{\rho}
\def\t{\tau}
\def\s{\sigma}
\def\z{\zeta}
\def\x{\xi}
\def\V={{{\bf\rm{V}}}}
 \def\A{{\cal{A}}}
 \def\B{{\cal{B}}}
 \def\C{{\cal{C}}}
 \def\D{{\cal{D}}}
\def\G{\Gamma}
\def\K{{\cal{K}}}
\def\O{\Omega}
\def\R{\bar{R}}
\def\T{{\cal{T}}}
\def\L{\Lambda}
\def\f{E_{\tau,\eta}(sl_2)}
\def\E{E_{\tau,\eta}(sl_n)}
\def\Zb{\mathbb{Z}}
\def\Cb{\mathbb{C}}

\def\R{\overline{R}}

\def\beq{\begin{equation}}
\def\eeq{\end{equation}}
\def\bea{\begin{eqnarray}}
\def\eea{\end{eqnarray}}
\def\ba{\begin{array}}
\def\ea{\end{array}}
\def\no{\nonumber}
\def\le{\langle}
\def\re{\rangle}
\def\lt{\left}
\def\rt{\right}

\newtheorem{Theorem}{Theorem}
\newtheorem{Definition}{Definition}
\newtheorem{Proposition}{Proposition}
\newtheorem{Lemma}{Lemma}
\newtheorem{Corollary}{Corollary}
\newcommand{\proof}[1]{{\bf Proof. }
        #1\begin{flushright}$\Box$\end{flushright}}

\baselineskip=20pt

\newfont{\elevenmib}{cmmib10 scaled\magstep1}
\newcommand{\preprint}{
   \begin{flushleft}
   \end{flushleft}\vspace{-1.3cm}
   \begin{flushright}\normalsize
   \end{flushright}}
\newcommand{\Title}[1]{{\baselineskip=26pt
   \begin{center} \Large \bf #1 \\ \ \\ \end{center}}}
\newcommand{\Author}{\begin{center}
   \large \bf
Yi Qiao${}^{a,b,c,d}$,~Pei Sun${}^{b,c,d,e,\,1}$,~Xiaotian Xu${}^{a,b,c,d}$, ~Junpeng Cao${}^{c,f,g\,1}$,~Guang-Liang Li${}^{h,\,1}$,~Wen-Li Yang${}^{a,b,c,d,\,1}$,
 ~Kangjie Shi${}^a$, and~Yupeng Wang${}^{f,\,}\footnote{Corresponding authors}$

 \end{center}}

\newcommand{\Address}{\begin{center}

     ${}^a$ Institute of Modern Physics, Northwest University,
     Xian 710127, China\\
     ${}^b$ Shaanxi Key Laboratory for Theoretical Physics Frontiers,  Xian 710127, China\\
     ${}^c$ Peng Huanwu Center for Fundamental Theory, Xian 710127, China\\
     ${}^d$ Fundamental Discipline Research Center for Quantum Science and Technology of Shaanxi Province, Xian 710127, China\\
     ${}^e$ School of Physics, Northwest University,  Xian 710127, China\\
     ${}^f$ Beijing National Laboratory for Condensed Matter
           Physics, Institute of Physics, Chinese Academy of Sciences, Beijing
           100190, China\\
     ${}^g$School of Physical Sciences, University of Chinese Academy of Sciences, Beijing 100049, China\\
     ${}^h$ Department of Applied Physics, Xian Jiaotong University, Xian 710049, China\\
     E-mails: qiaoyi\_joy@foxmail.com,\, sunpei@nwu.edu.cn,\, xtxu@nwu.edu.cn,\, junpengcao@iphy.ac.cn,\, leegl@mail.xjtu.edu.cn,\, wlyang@nwu.edu.cn,\, kjshi@nwu.edu.cn,\, yupeng@iphy.ac.cn

   \end{center}}
\newcommand{\Accepted}[1]{\begin{center}
   {\large \sf #1}\\ \vspace{1mm}{\small \sf Accepted for Publication}
   \end{center}}

\preprint
\thispagestyle{empty}
\bigskip

\Title{An elliptic generalization of the $D^{(2)}_{2}$ Spin chain } \Author

\Address

\begin{abstract}
An integrable elliptic generalization of the closed $D^{(2)}_2$ quantum spin chain is constructed. Taking the trigonometric limit, we have shown that the corresponding R-matrix reduces to that
of the $D^{(2)}_2$ one, up to a similarity transformation. Basing on the very fact of the factorization for the R-matrix and
the corresponding transfer matrix which indeed ensures the integrability of the model, we give the exact solution of the model which can be expressed in terms of
a homogeneous T-Q relation.

\vspace{0.18truecm}

\end{abstract}
\newpage

\section{Introduction}
\label{intro} \setcounter{equation}{0}

The antiferromagnetic (AF) Potts model has been the subject of intense studies for many years \cite{Pot52, Bax82, bax2, Sal91, Jac06, Rob20, Rob21, Jac26} due to its unusual feature that the conformal field
theory (CFT) describing its continuum limit is "non-compact" and leading to the observation of continuum of critical exponents \cite{IKh08, Can13, Fra14, Baz19, Jac26}, which also shares the striking feature with
the staggered six-vertex model \cite{Tem71, Bax76, Bax82, Sal91, Jac06, IKh08, Fra14, Baz19, Rob20, Rob21}. Interestingly, it has been
shown in \cite{Rob20} that the antiferromagnetic Potts model and the staggered trigonometric six-vertex model (or the XXZ spin-$\frac{1}{2}$ chain) are related to the $D^{(2)}_2$ model (an integrable model associated with the twisted affine algebra $D^{(2)}_2$, which serves as the simplest and fundamental integrable model among the $D^{(2)}_n$ models \cite{Baz85, Jim86, Baz87}). Moreover, it has been further shown \cite{Rob21} that an open $D^{(2)}_2$
spin chain with a particular integrable boundary condition is closely related to the $SL(2, \mathbb{R})/U(1)$ Euclidean black hole and the string theory \cite{Wit91, Dij92, Mal01, Mal01-1, Han02, Baz20}.
These significant findings of the correspondence  have paved the way to studying the associated problems by the Bethe ansatz method. Therefore, the $D^{(2)}_2$ spin chain with some integrable boundary conditions then has
been intensively investigated \cite{Mar99, Mar00, Mal06, Nep17, Nep18, Nep19, Baz21, Nep21, Li22}.

It is well known that the eight-vertex model (or the XYZ spin-$\frac 12$ chain) is an integrable elliptic (anisotropic) generalization of the six-vertex model and itself is an important a typical model in statistical
physics, one-dimensional magnetism \cite{bax2, 2-baxter1, 2-baxter1-1, xyz2, yb, takahashi, Wan15}. The exact solution has been intensively studied \cite{xyz2, yb, takahashi, Wan15, Fan96, Yan06, Cao14}.  The aim of this paper is to construct such an integrable elliptic generalization of the $D^{(2)}_2$ model, which is equivalent to the original $D^{(2)}_2$ one under the trigonometric limit.  Following the method proposed in \cite{Cao1, Wan15}, we shall give the exact solution of the closed elliptic $D^{(2)}_2$ spin chain (or with the periodic condition).

The paper is organized as follows.  Section 2 serves as an introduction of our notations and some basic ingredients. After constructing the integrable elliptic generalization of the $D^{(2)}_2$ model by giving
the associated R-matrix $\mathbb{R}(u)$, we list some properties of the R-matrix which enable one to construct the local Hamiltonian with the nearest neighbour (NN) local interaction under the periodic boundary condition.
We also give its relation to the $D^{(2)}_2$ R-matrix after taking the trigonometric limit. In
Section 3, the factorization of the R-matrix $\mathbb{R}(u)$ is presented, which plays an important role to derive the product form of the associated transfer matrices.
Section 4, basing on the very fact of the factorization for the R-matrix and
the corresponding transfer matrix, we give the exact solution of the model which can be expressed in terms of
a homogeneous T-Q relation. Some detailed technique derivations are presented in the Appendices A, B $\&$ C.


\section{ Transfer matrix}
\label{XYZ} \setcounter{equation}{0}

Let us fix $\tau$ such that ${\rm Im}(\tau)>0$ and a generic
complex number $\eta$. Introduce the elliptic functions
\begin{eqnarray} &&\theta\left[
\begin{array}{c}
a\\b
\end{array}\right](u,\tau)=\sum_{m=-\infty}^{\infty}
\exp\left\{i\pi\left[(m+a)^2\tau+2(m+a)(u+b)\right]\right\},\label{Function-a-b}\\
&&\s(u)=\theta\left[\begin{array}{c}\frac{1}{2}\\[2pt]\frac{1}{2}
\end{array}\right](u,\tau),\quad \zeta(u)=\frac{\partial}{\partial u}
\left\{\ln\s(u)\right\}.\label{Function}
\end{eqnarray}

Let $\mathbb{V}$ be a $4$-dimensional vector space. We introduce an R-matrix $\mathbb{R}(u)\in {\rm
End}(\mathbb{V}\otimes \mathbb{V})$, which is an elliptic generalization of that associated with the quantum twisted affine  $D^{(2)}_2$ algebra \cite{Rob20}, given by  \bea
\resizebox{\textwidth}{!}{$
\mathbb{R}(u)=\Omega(u)\lt(\begin{array}{cccc|cccc|cccc|cccc}r_{1} & 0 & 0 & r_{2} & 0 & r_{3} & r_{4} & 0 & 0 & r_{5} & r_{3} & 0 & r_{2} & 0 & 0 & r_{6}\\
	0 & r_{7} & 0 & 0 & r_{8} & 0 & 0 & r_{9} & -r_{10} & 0 & 0 & r_{11} & 0 & 0 & 0 & 0\\
	0 & 0 & r_{7} & 0 & r_{10} & 0 & 0 & r_{12} & r_{13} & 0 & 0 & -r_{9} & 0 & 0 & 0 & 0\\
	-r_{2} & 0 & 0 & r_{14} & 0 & r_{15} & r_{16} & 0 & 0 & r_{16} & r_{17} & 0 & r_{18} & 0 & 0 & -r_{2}\\\hline
	0 & r_{13} & r_{10} & 0 & r_{7} & 0 & 0 & 0 & 0 & 0 & 0 & 0 & 0 & -r_{9} & r_{12} & 0\\
	r_{3} & 0 & 0 & -r_{17} & 0 & r_{19} & r_{20} & 0 & 0 & r_{20} & r_{21} & 0 & -r_{15} & 0 & 0 & r_{3}\\
	-r_{5} & 0 & 0 & r_{16} & 0 & -r_{20} & r_{22} & 0 & 0 & r_{23} & -r_{20} & 0 & r_{16} & 0 & 0 & -r_{4}\\
	0 & r_{9} & r_{11} & 0 & 0 & 0 & 0 & r_{7} & 0 & 0 & 0 & 0 & 0 & r_{8} & -r_{10} & 0\\\hline
	0 & -r_{10} & r_{8} & 0 & 0 & 0 & 0 & 0 & r_{7} & 0 & 0 & 0 & 0 & r_{11} & r_{9} & 0\\
	-r_{4} & 0 & 0 & r_{16} & 0 & -r_{20} & r_{23} & 0 & 0 & r_{22} & -r_{20} & 0 & r_{16} & 0 & 0 & -r_{5}\\
	r_{3} & 0 & 0 & -r_{15} & 0 & r_{21} & r_{20} & 0 & 0 & r_{20} & r_{19} & 0 & -r_{17} & 0 & 0 & r_{3}\\
	0 & r_{12} & -r_{9} & 0 & 0 & 0 & 0 & 0 & 0 & 0 & 0 & r_{7} & 0 & r_{10} & r_{13} & 0\\\hline
	-r_{2} & 0 & 0 & r_{18} & 0 & r_{17} & r_{16} & 0 & 0 & r_{16} & r_{15} & 0 & r_{14} & 0 & 0 & -r_{2}\\
	0 & 0 & 0 & 0 & -r_{9} & 0 & 0 & r_{13} & r_{12} & 0 & 0 & r_{10} & 0 & r_{7} & 0 & 0\\
	0 & 0 & 0 & 0 & r_{11} & 0 & 0 & -r_{10} & r_{9} & 0 & 0 & r_{8} & 0 & 0 & r_{7} & 0\\
	r_{6} & 0 & 0 & r_{2} & 0 & r_{3} & r_{5} & 0 & 0 & r_{4} & r_{3} & 0 & r_{2} & 0 & 0 & r_{1}
\end{array}\rt).
$} \nonumber\\
\label{r-matrix}\eea The
non-vanishing matrix elements  are given by

\begin{eqnarray}
r_{1}(u)&=&-\beta(z)^2[\beta(y)-\beta(\eta)^2\beta(x)][\beta(y)+\beta(\eta)^2\beta(x)], \quad
r_{2}(u)=\beta(\eta)^2\beta(u)\beta(z)\beta_1(y,x),\nonumber\\
r_{3}(u)&=&\beta(\eta)\beta(u)\beta(x)\beta(z)\beta_1(\eta,y), \quad
r_{4}(u)=-\beta(\eta)\beta(x)\beta(z)^2\beta_1(\eta,y),\nonumber\\
r_{5}(u)&=&-\beta(\eta)\beta(u)^2\beta(x)\beta(z)^2\beta_2(\eta,y), \quad
r_{6}(u)=-\beta(\eta)^2\beta(u)^2\beta(z)^2\beta_1(y,x),\nonumber\\
r_{7}(u)&=&-\beta(u)\beta(y)\beta(x)\beta(z)\beta_2(\eta,\eta), \quad
r_{8}(u)=\beta(\eta)\beta(y)\beta_1(\eta,x),
\nonumber\\
r_{9}(u)&=&-\beta(\eta)\beta(u)\beta(y)\beta(z)\beta_1(\eta,x), \quad
r_{10}(u)=\beta(\eta)\beta(u)\beta(y)\beta(z)\beta_2(\eta,x),\nonumber\\
r_{11}(u)&=&\beta(\eta)\beta(u)^2\beta(y)\beta_1(\eta,x), \quad
r_{12}(u)=-\beta(\eta)\beta(u)^2\beta(y)\beta(z)^2\beta_2(\eta,x),\nonumber\\
r_{13}(u)&=&-\beta(\eta)\beta(y)\beta(z)^2\beta_2(\eta,x), \quad
r_{14}(u)=\beta(u)^2[\beta(\eta)^2\beta(y)-\beta(x)][\beta(\eta)^2\beta(y)+\beta(x)], \nonumber\\
r_{15}(u)&=&\beta(\eta)\beta(u)^2\beta(x)\beta_2(\eta,y), \quad
r_{16}(u)=-\beta(\eta)\beta(u)\beta(x)\beta(z)\beta_2(\eta,y),\nonumber \\
r_{17}(u)&=&\beta(\eta)\beta(x)\beta_1(\eta,y),
\quad
r_{18}(u)=\beta(\eta)^2\beta_1(y,x),\nonumber\\
r_{19}(u)&=&[\beta(\eta)^2-\beta(y)\beta(x)][\beta(\eta)^2+\beta(y)\beta(x)],\quad
r_{20}(u)=-\beta(\eta)^2\beta(u)\beta(z)\beta_2(y,x), \nonumber\\
r_{21}(u)&=&\beta(\eta)^2\beta(u)^2\beta_2(y,x),
\quad
r_{22}(u)=\beta(u)^2\beta(z)^2[\beta(\eta)^2\beta(y)\beta(x)-1][\beta(\eta)^2\beta(y)\beta(x)+1],
\nonumber\\
r_{23}(u)&=&-\beta(\eta)^2\beta(z)^2\beta_2(y,x),\label{r-func}\end{eqnarray}
where the normalization factor $\Omega(u)$ is given by
\begin{equation}
	\Omega(u)=\frac{[\theta_0(u)\theta_0(y)\theta_0(x)\theta_0(z)]^2}{\theta_0(0)^4\theta_1(\eta)^4}.
\end{equation}
For compactness, we have introduced
$x=u+\frac{1}{2}$, $y=u+\eta$,
$z=u+\eta+\frac{1}{2}$, and  $\b(u)$, $\b_i(u)$ in (\ref{r-func}) given by
\begin{equation}
	\begin{aligned}
		\beta_1(v,w)
		&= \frac{\theta_0(0)^2\theta_1(v+w)\theta_1(v-w)}
		{\theta_0(v)^2\theta_0(w)^2}
		= \beta(v)^2-\beta(w)^2,\\[6pt]
		\beta_2(v,w)
		&= \frac{\theta_0(0)^2\theta_0(v+w)\theta_0(v-w)}
		{\theta_0(v)^2\theta_0(w)^2}
		= 1-\beta(v)^2\beta(w)^2,
	\end{aligned}
\end{equation}
with
\begin{equation}
	\begin{gathered}
		\theta_0(v)=\theta
		\begin{bmatrix}
			0\\ \frac{1}{2}
		\end{bmatrix}(v,2\tau),\quad
		\theta_1(v)=\theta
		\begin{bmatrix}
			\frac{1}{2}\\ \frac{1}{2}
		\end{bmatrix}(v,2\tau),\\
		\beta(v)=\frac{\theta_1(v)}{\theta_0(v)}.
	\end{gathered}
\end{equation}
Here $u$ is the spectral parameter and $\eta$ is the so-called
crossing parameter. The R-matrix given by (\ref{r-matrix})-(\ref{r-func}) satisfies the quantum Yang-Baxter equation (QYBE) \cite{bax2, yang1, yang2}
\begin{eqnarray}
\mathbb{R}_{\hollow{1}\, \hollow{2}}(u_1-u_2)\,\mathbb{R}_{\hollow{1}\,\hollow{3}}(u_1-u_3)\,\mathbb{R}_{\hollow{2}\,\hollow{3}}(u_2-u_3)=
\mathbb{R}_{\hollow{2}\,\hollow{3}}(u_2-u_3)\,\mathbb{R}_{\hollow{1}\,\hollow{3}}(u_1-u_3)\,\mathbb{R}_{\hollow{1}\,\hollow{2}}(u_1-u_2),\label{QYB}
\end{eqnarray}
In addition to the QYBE, the R-matrix also satisfies the properties:
\bea
&&\hspace{-1.5cm}\mbox{ Initial
condition}:\,\mathbb{R}_{\hollow{1}\,\hollow{2}}(0)=\Theta_0\, \mathbb{P}_{\hollow{1}\,\hollow{2}},\quad \Theta_0=\rho_1(\frac{1}{2}),\label{Int-R}\\
&&\hspace{-1.5cm}\mbox{ Unitarity
relation}:\,\mathbb{R}_{\hollow{1}\,\hollow{2}}(u)\mathbb{R}_{\hollow{2}\,\hollow{1}}(-u)= \Theta_1(u)\times\,{\rm id},
\quad \Theta_1(u)=\rho_1^2(u-\frac{1}{2})\,\rho^2_1(u),\label{Unitarity}\\
&&\hspace{-1.5cm}\mbox{ Crossing-
Unitarity}:\,\mathbb{R}^{t_{\hollow{2}}}_{\hollow{1}\,\hollow{2}}(u)\,\mathbb{R}^{t_{\hollow{2}}}_{\hollow{2}\,\hollow{1}}(-u-2\eta)=\Theta_2(u)\times\,{\rm id},\,\Theta_2(u)=\rho_2^2(u-\frac{1}{2})\,\rho^2_2(u),
\label{crosing-unitarity}
\eea
where the functions $\rho_1(u)$ and $\rho_2(u)$ are respectively given by (\ref{Unitarity-B}) and (\ref{crosing-unitarity-B}) later,
and $\mathbb{R}_{\hollow{2}\,\hollow{1}}(u)=\mathbb{P}_{\hollow{1}\,\hollow{2}}\mathbb{R}_{\hollow{1}\,\hollow{2}}(u)\mathbb{P}_{\hollow{1}\,\hollow{2}}$ with $\mathbb{P}_{\hollow{1}\,\hollow{2}}$ being
the usual permutation operator and $t_{\hollow{i}}$ denotes transposition in
the $\hollow{i}$-th space. Throughout this paper we adopt the standard notations: for
any matrix $\mathbb{A}\in {\rm End}(\mathbb{V})$, $\mathbb{A}_{\hollow{j}}$ is an embedding operator
in the tensor space $\mathbb{V}\otimes \mathbb{V}\otimes\cdots\otimes\mathbb{V}$, which acts
as $\mathbb{A}$ on the $\hollow{j}$-th space and as identity on the other factor
spaces; $\mathbb{R}_{\hollow{i}\,\hollow{j}}(u)$ is an embedding operator of R-matrix in the
tensor space, which acts as identity on the factor spaces except
for the $\hollow{i}$-th and $\hollow{j}$-th ones.

After taking the trigonometric limit (i.e. $\tau\rightarrow +i\infty$) and redefining the parameters as
\bea
u\longrightarrow \frac{iu}{2\pi},\quad\quad \eta\longrightarrow 1-\frac{i\eta}{\pi},\label{Repar}
\eea
the resulting  R-matrix $\mathbb{R}(u)$ of (\ref{r-matrix})-(\ref{r-func}) become the R-matrix (Here let us denote it by $\mathbb{R}^{D^{(2)}_2}_{\hollow{1}\,\hollow{2}}(u)$, we shall use the definition \cite{Nep21}) associated with the $D^{(2)}_2$ algebra constructed in \cite{Jim86, Mar99, Fra12, Rob20, Nep21}, up to a similarity transformation. Namely,
\bea
&&\tilde{\mathbb{R}}_{\hollow{1}\,\hollow{2}}(u)\stackrel{\rm def}{=}\lt.\mathbb{R}_{\hollow{1}\,\hollow{2}}(u)\rt|_{u\longrightarrow \frac{iu}{2\pi},\,\eta\longrightarrow 1-\frac{i\eta}{\pi}},\nonumber\\[6pt]
&&\mathbb{R}^{D^{(2)}_2}_{\hollow{1}\,\hollow{2}}(u)=(2\sinh\eta)^4\tilde{B}_{1\bar{1}}(u)\,\tilde{B}_{2\bar{2}}(0)\lt(\lim_{\tau\rightarrow +i\infty}\tilde{\mathbb{R}}_{\hollow{1}\,\hollow{2}}(u)\rt)\tilde{B}^{-1}_{1\bar{1}}(u)\,\tilde{B}^{-1}_{2\bar{2}}(0),\label{R-matrix-relation}
\eea
where the similarity transformation $\tilde{B}(u)$ is
\bea
\tilde{B}(u)= \lt(\begin{array}{llll}e^{-\frac{u}{2}}&&&\\[6pt]
&\frac{e^{-\frac{i\pi}{4}}\cosh\frac{\eta}{2}}{\sqrt{\cosh\eta}}&-\frac{e^{\frac{i\pi}{4}}\sinh\frac{\eta}{2}}{\sqrt{\cosh\eta}}&\\[6pt]
&-\frac{e^{-\frac{i\pi}{4}}\sinh\frac{\eta}{2}}{\sqrt{\cosh\eta}}&-\frac{e^{\frac{i\pi}{4}}\cosh\frac{\eta}{2}}{\sqrt{\cosh\eta}}&\\[6pt]
&&&e^{\frac{u}{2}}
\end{array}\rt).
\eea
We shall give the detailed proof of the relation (\ref{R-matrix-relation}) in Appendix B.

Let us introduce the monodromy matrix associated with the elliptic $D^{(2)}_2$ R-matrix $\mathbb{R}(u)$ given by (\ref{r-matrix})-(\ref{r-func})
\begin{eqnarray}
\mathbb{T}_{\hollow{0}}(u)&=&\mathbb{R}_{\hollow{0}\,\hollow{N}}(u)\cdots\mathbb{R}_{\hollow{0}\,\hollow{1}}(u),\label{Mon-1}
\end{eqnarray}
and the associated
transfer matrix $\tau(u)$ of the elliptic $D^{(2)}_2$ spin chain with the periodic boundary condition (or the closed elliptic $D^{(2)}_2$ spin chain ) given by \cite{bax2}
 \begin{eqnarray}
 \tau(u)=tr_{\hollow{0}}\lt\{\mathbb{T}_{\hollow{0}}(u)\rt\},\label{trans-per}
 \end{eqnarray} $tr_{\hollow{0}}$ denotes trace over the
``auxiliary space" $\hollow{0}$. The Hamiltonian with the NN local interaction under the periodic boundary condition can be constructed \cite{yb}
\begin{eqnarray}
\hspace{-0.16truecm}H=\Theta_0\hspace{-0.16truecm}\lt.\frac{\partial}{\partial u}\ln\tau(u)\rt|_{u=0}
\hspace{-0.16truecm}=\hspace{-0.26truecm}\sum_{\hollow{j}=1}^{N}h_{\hollow{j},\,\hollow{j+1}},\,
h_{\hollow{j},\,\hollow{j+1}}\hspace{-0.16truecm}=\hspace{-0.12truecm}
\frac{\partial}{\partial u}\lt.\check{\mathbb{R}}_{\hollow{j},\,\hollow{j+1}}(u)\right|_{u=0}
\hspace{-0.16truecm}=\hspace{-0.16truecm}\frac{\partial}{\partial u}\lt\{\mathbb{P}_{\hollow{j},\,\hollow{j+1}}\mathbb{R}_{\hollow{j},\,\hollow{j+1}}(u)\right\}\lt.\right|_{u=0},\label{ham}
\end{eqnarray}
where the constant $\Theta_0$ is given in (\ref{Int-R}). The periodic boundary condition reads
\bea
h_{\hollow{N},\,\hollow{N+1}}=h_{\hollow{N},\,\hollow{1}}.\label{PBC}
\eea
The QYBE (\ref{QYB})  leads to
the fact that the transfer matrices with different spectral
parameters commute with each other \cite{yb}: $[\tau(u),\,\tau(v)]=0$.
Then $\tau(u)$ serves as the generating functional of the conserved
quantities of the corresponding system, which ensures the
integrability of the closed elliptic $D^{(2)}_2$ spin chain  described by the Hamiltonian  with the NN local interaction (\ref{ham}) and the periodic boundary condition
 (\ref{PBC}). The aim of this paper focuses on the exact solution of eigenvalues of the transfer matrix (\ref{trans-per}) (including the Hamiltonian (\ref{ham})-(\ref{PBC})).


\section{Factorization of the R-matrix }
\label{Factor}
\setcounter{equation}{0}
Similar as the fact that the $D^{(2)}_2$ R-matrix \cite{Jim86} has been shown \cite{Mar99, Fra12, Rob20, Nep21} to be of the product form  of the trigonometric six-vertex R-matrix of the type \cite{Zvy01}  up to a similarity transformation, our elliptic R-matrix $\mathbb{R}(u)$ given by (\ref{r-matrix}) can be given in terms of the product form of the well-known eight-vertex model R-matrix. In order to exploit the factorization, it is very
convenient to replace each index $\hollow{j}$ in (\ref{Mon-1}) (which corresponds to a 4-dimensional vector space) by a pair of indices $(j, \bar{j})$ (each of which corresponds to a 2-dimensional vector space $\Cb^2$). Then the R-matrix $\mathbb{R}_{\hollow{1}\,\hollow{2}}(u)$ can be expressed in terms of the product form as
\bea
\mathbb{R}_{\hollow{1}\,\hollow{2}}(u)\equiv \mathbb{R}_{1\bar{1},\, 2\bar{2}}(u)=
R_{1\,\bar{2}}(u+\frac{1}{2})\,R_{1\,2}(u)\,R_{\bar{1}\,\bar{2}}(u)\,
R_{\bar{1}\,2}(u-\frac{1}{2}),\label{R-Factor}
\eea
where the eight-vertex model R-matrix $R(u)\in {\rm
End}(\Cb^2\otimes \Cb^2)$ is given by  \bea
R(u)=\lt(\begin{array}{llll}\bar{a}(u)&&&\bar{d}(u)\\&\bar{b}(u)&\bar{c}(u)&\\
&\bar{c}(u)&\bar{b}(u)&\\\bar{d}(u)&&&\bar{a}(u)\end{array}\rt). \label{r-matrix-B}\eea The
non-vanishing matrix elements  are \cite{bax2}\bea
&&\hspace{-2.0truecm}\bar{a}(u)\hspace{-0.1truecm}=
\hspace{-0.1truecm}\frac{\theta\lt[\begin{array}{c} 0\\\frac{1}{2}
\end{array}\rt]\hspace{-0.16truecm}(u,2\tau)\hspace{0.12truecm}
\theta\lt[\begin{array}{c} \frac{1}{2}\\[2pt]\frac{1}{2}
\end{array}\rt]\hspace{-0.16truecm}(u+\eta,2\tau)}{\theta\lt[\begin{array}{c} 0\\\frac{1}{2}
\end{array}\rt]\hspace{-0.16truecm}(0,2\tau)\hspace{0.12truecm}
\theta\lt[\begin{array}{c} \frac{1}{2}\\[2pt]\frac{1}{2}
\end{array}\rt]\hspace{-0.16truecm}(\eta,2\tau)},\quad
\bar{b}(u)\hspace{-0.1truecm}=\hspace{-0.1truecm}\frac{\theta\lt[\begin{array}{c}
\frac{1}{2}\\[2pt]\frac{1}{2}
\end{array}\rt]\hspace{-0.16truecm}(u,2\tau)\hspace{0.12truecm}
 \theta\lt[\begin{array}{c} 0\\\frac{1}{2}
\end{array}\rt]\hspace{-0.16truecm}(u+\eta,2\tau)}
{\theta\lt[\begin{array}{c} 0\\\frac{1}{2}
\end{array}\rt]\hspace{-0.16truecm}(0,2\tau)\hspace{0.12truecm}
\theta\lt[\begin{array}{c} \frac{1}{2}\\[2pt]\frac{1}{2}
\end{array}\rt]\hspace{-0.16truecm}(\eta,2\tau)},\no\\
&&\hspace{-2.0truecm}\bar{c}(u)\hspace{-0.1truecm}=
\hspace{-0.1truecm}\frac{\theta\lt[\begin{array}{c} 0\\\frac{1}{2}
\end{array}\rt]\hspace{-0.16truecm}(u,2\tau)\hspace{0.12truecm}
 \theta\lt[\begin{array}{c} 0\\\frac{1}{2}
\end{array}\rt]\hspace{-0.16truecm}(u+\eta,2\tau)}
{\theta\lt[\begin{array}{c} 0\\\frac{1}{2}
\end{array}\rt]\hspace{-0.16truecm}(0,2\tau)\hspace{0.12truecm}
\theta\lt[\begin{array}{c} 0\\\frac{1}{2}
\end{array}\rt]\hspace{-0.16truecm}(\eta,2\tau)},\quad
\bar{d}(u)\hspace{-0.1truecm}=\hspace{-0.1truecm}\frac{\theta\lt[\begin{array}{c}
\frac{1}{2}\\[2pt]\frac{1}{2}
\end{array}\rt]\hspace{-0.16truecm}(u,2\tau)\hspace{0.12truecm}
 \theta\lt[\begin{array}{c} \frac{1}{2}\\[2pt]\frac{1}{2}
\end{array}\rt]\hspace{-0.16truecm}(u+\eta,2\tau)}
{\theta\lt[\begin{array}{c} 0\\\frac{1}{2}
\end{array}\rt]\hspace{-0.16truecm}(0,2\tau)\hspace{0.12truecm}
\theta\lt[\begin{array}{c} 0\\\frac{1}{2}
\end{array}\rt]\hspace{-0.16truecm}(\eta,2\tau)}.\label{r-func-B}
\eea
In deriving the relation (\ref{R-Factor}), we have used the identities of the elliptic functions listed in Appendix A.
In addition to  the quantum Yang-Baxter
equation (QYBE),
\begin{eqnarray}
R_{12}(u_1-u_2)R_{13}(u_1-u_3)R_{23}(u_2-u_3)=R_{23}(u_2-u_3)R_{13}(u_1-u_3)R_{12}(u_1-u_2),\label{QYB-B}
\end{eqnarray}
the R-matrix also satisfies the following properties:
\bea
&&\hspace{-1.5cm}\mbox{ Initial
condition}:\,R_{12}(0)= P_{12},\label{Int-R-B}\\
&&\hspace{-1.5cm}\mbox{ Unitarity
relation}:\,R_{12}(u)R_{21}(-u)= \rho_1(u)\times\,{\rm id},
\quad \rho_1(u)=-\frac{\s(u-\eta)\s(u+\eta)}{\s(\eta)\,\s(\eta)},\label{Unitarity-B}\\
&&\hspace{-1.5cm}\mbox{ Crossing
relation}:\,R_{12}(u)=V_1R_{12}^{t_2}(-u-\eta)V_1,\quad
V=-i\s^y,
\label{crossing-relation-B}\\
&&\hspace{-1.5cm}\mbox{ PT-symmetry}:\,R_{12}(u)=R_{21}(u)=R^{t_1\,t_2}_{12}(u),\label{PT}\\
&&\hspace{-1.5cm}\mbox{$Z_2$-symmetry}: \,
\qquad\s^i_1\s^i_2R_{1,2}(u)=R_{1,2}(u)\s^i_1\s^i_2,\quad
\mbox{for}\,\,
i=x,y,z.\label{Z2-sym}
\eea
Here $\{\s^i|i=x,y,z\}$ are the Pauli matrices, and $R_{21}(u)=P_{12}R_{12}(u)P_{12}$ with $P_{12}$ being
the usual permutation operator in the tensor space $\Cb^2\otimes \Cb^2$ (c.f. $\mathbb{P}_{\hollow{1}\,\hollow{2}}$, see (\ref{Int-R}) in the previous section. Moreover, the relation (\ref{R-Factor}) implies that $\mathbb{P}_{\hollow{1}\,\hollow{2}}=P_{1\,2}\,P_{\bar{1}\,\bar{2}} $) and $t_i$ denotes transposition in
the $i$-th space. The unitarity (\ref{Unitarity-B}) and crossing relation (\ref{crossing-relation-B}) lead to
the crossing-unitarity relation
\bea
R_{12}^{t_2}(u)\, R_{21}^{t_2}(-u-2\eta)=\rho_2(u)\times {\rm id},\quad
\rho_2(u)=-\frac{\s(u)\s(u+2\eta)}{\s(\eta)\,\s(\eta)}.\label{crosing-unitarity-B}
\eea
Here we also adopt the similar notations as those used in the tensor spaces of $\mathbb{V}$: for
any matrix $A\in {\rm End}(\Cb^2)$, $A_j$ is an embedding operator
in the tensor space $\Cb^2\otimes \Cb^2\otimes\cdots$, which acts
as $A$ on the $j$-th space and as identity on the other factor
spaces; $R_{i\,j}(u)$ is an embedding operator of R-matrix in the
tensor space, which acts as identity on the factor spaces except
for the $i$-th and $j$-th ones.

It is easy to check that the QYBE (\ref{QYB-B}) and the properties (\ref{Int-R-B}), (\ref{Unitarity-B}) and (\ref{crosing-unitarity-B}) of the eight-vertex R-matrix immediately give rise to
those (\ref{QYB})-(\ref{crosing-unitarity}) of
the elliptic $D^{(2)}_{2}$ R-matrix $\mathbb{R}(u)$ given by (\ref{r-matrix})-(\ref{r-func}).


\section{Exact solution of the closed elliptic $D^{(2)}_2$ spin chain }
\label{Exact solution}
\setcounter{equation}{0}
Thanks to the product form (\ref{R-Factor}) of the R-matrix $\mathbb{R}(u)$, the monodromy matrix for the elliptic $D^{(2)}_2$ chain of length $N$ defined by (\ref{Mon-1})
can be expressed as
\bea
\mathbb{T}_{\hollow{0}}(u)\hspace{-0.46truecm}&\equiv & \hspace{-0.46truecm}\mathbb{T}_{0\,\bar{0}}(u)=
\mathbb{R}_{0\bar{0},\,N\bar{N}}(u)\cdots\mathbb{R}_{0\bar{0},\,1\bar{1}}(u)\nonumber\\
&\stackrel{(\ref{R-Factor})}{=}&\hspace{-0.26truecm}\lt(\hspace{-0.12truecm}R_{0\bar{N}}(u\hspace{-0.12truecm}+\hspace{-0.12truecm}\frac{1}{2})R_{0N}(u)R_{\bar{0}\bar{N}}(u)
R_{\bar{0}N}(u\hspace{-0.12truecm}-\hspace{-0.12truecm}\frac{1}{2})\hspace{-0.12truecm}\rt)\hspace{-0.12truecm}\cdots
\hspace{-0.12truecm}\lt(\hspace{-0.12truecm}R_{0\bar{1}}(u\hspace{-0.12truecm}+\hspace{-0.12truecm}\frac{1}{2})R_{01}(u)R_{\bar{0}\bar{1}}(u)
R_{\bar{0}1}(u\hspace{-0.12truecm}-\hspace{-0.12truecm}\frac{1}{2})\hspace{-0.12truecm}\rt)\nonumber\\[6pt]
&=&\hspace{-0.26truecm}\lt(\hspace{-0.12truecm}R_{0\bar{N}}(u+\frac{1}{4}+\frac{1}{4})\,R_{0N}(u+\frac{1}{4}-\frac{1}{4})\hspace{-0.12truecm}\rt)\cdots
\hspace{-0.12truecm}\lt(\hspace{-0.12truecm}R_{0\bar{1}}(u+\frac{1}{4}+\frac{1}{4})\,R_{01}(u+\frac{1}{4}-\frac{1}{4})\hspace{-0.12truecm}\rt)\nonumber\\[6pt]
&&\quad\times \lt(\hspace{-0.12truecm}R_{\bar{0}\bar{N}}(u\hspace{-0.12truecm}-\hspace{-0.12truecm}\frac{1}{4}\hspace{-0.12truecm}+\hspace{-0.12truecm}\frac{1}{4})\,
R_{\bar{0}N}(u\hspace{-0.12truecm}-\hspace{-0.12truecm}\frac{1}{4}\hspace{-0.12truecm}-\hspace{-0.12truecm}\frac{1}{4})\hspace{-0.12truecm}\rt)\hspace{-0.12truecm}\cdots
\hspace{-0.12truecm}\lt(\hspace{-0.12truecm}R_{\bar{0}\bar{1}}(u\hspace{-0.12truecm}-\hspace{-0.12truecm}\frac{1}{4}\hspace{-0.12truecm}+\hspace{-0.12truecm}\frac{1}{4})\,
R_{\bar{0}1}(u\hspace{-0.12truecm}-\hspace{-0.12truecm}\frac{1}{4}\hspace{-0.12truecm}-\hspace{-0.12truecm}\frac{1}{4})\hspace{-0.12truecm}\rt)\nonumber\\[6pt]
&\stackrel{{\rm def}}{=}&T_{0}(u+\frac{1}{4};\{\frac{1}{4},-\frac{1}{4}\})\,\,T_{\bar{0}}(u-\frac{1}{4};\{\frac{1}{4},-\frac{1}{4}\}),\label{Mon-2}
\eea
where the  resulting monodromy matrix $T_{i}(u;\{\frac{1}{4},-\frac{1}{4}\})$ (with $i=0,\,\bar{0}$) of the staggered closed eight-vertex chain of length $2N$ is defined by
\bea
T_{i}(u;\{\frac{1}{4},-\frac{1}{4}\})=\lt(R_{i\,\bar{N}}(u\hspace{-0.12truecm}+\hspace{-0.12truecm}\frac{1}{4})\,
R_{i\,N}(u\hspace{-0.12truecm}-\hspace{-0.12truecm}\frac{1}{4})\rt)\cdots
\lt(R_{i\,\bar{1}}(u\hspace{-0.12truecm}+\hspace{-0.12truecm}\frac{1}{4})\,
R_{i\,1}(u\hspace{-0.12truecm}-\hspace{-0.12truecm}\frac{1}{4})\rt),\quad i=0,\,\bar{0}.\label{Mon-3}
\eea
This implies that the transfer matrix $\tau(u)$ given by (\ref{trans-per}) of the closed elliptic $D^{(2)}_2$ can be re-expressed in terms of the product form as
\bea
&&\tau(u)=t(u+\frac{1}{4};\{\frac{1}{4},\,-\frac{1}{4}\})\,t(u-\frac{1}{4};\{\frac{1}{4},\,-\frac{1}{4}\}),\label{trans-per-factor}
\eea
where the transfer matrix of the staggered eight-vertex chain of length $2N$ is given by
\bea
t(u;\{\frac{1}{4},\,-\frac{1}{4}\})=
tr_{0}\lt\{T_{0}(u;\{\frac{1}{4},-\frac{1}{4}\})\rt\}.\label{trans-per-8}
\eea
Since the fact that the transfer matrices $\tau(u)$ and $t(u;\{\frac{1}{4},\,-\frac{1}{4}\})$ commute with each other with different spectrum parameters, namely,
\bea
[\tau(u),\,\tau(v)]=[\tau(u),\,t(v;\{\frac{1}{4},\,-\frac{1}{4}\})]=[t(u;\{\frac{1}{4},\,-\frac{1}{4}\}),\,t(v;\{\frac{1}{4},\,-\frac{1}{4}\})]=0,
\eea
they may share common eigenstates. Let $|\Phi\rangle$ be a common eigenstate, and $\Lambda(u)$, $\bar{\Lambda}(u)$ be the corresponding eigenvalues, namely,
\bea
\tau(u)\,|\Phi\rangle=\Lambda(u)\,|\Phi\rangle,\quad t(u;\{\frac{1}{4},\,-\frac{1}{4}\})\,|\Phi\rangle=\bar{\Lambda}(u)|\Phi\rangle.
\eea
The relation (\ref{trans-per-factor}) leads to that
\bea
\Lambda(u)=\bar{\Lambda}(u+\frac{1}{4})\,\bar{\Lambda}(u-\frac{1}{4}).\label{eigenvaules}
\eea

The expressions (\ref{Mon-3}) and (\ref{trans-per-8}) imply that the transfer matrix $t(u;\{\frac{1}{4},\,-\frac{1}{4}\})$ is nothing but the transfer matrix of
the inhomogeneous closed XYZ spin-$\frac{1}{2}$ chain with the length $2N$ and the staggered inhomogeneous parameters $\{\frac{1}{4},\,-\frac{1}{4}\}$. The associated eigenvalues can be expressed
in terms of a T-Q relation by the off-diagonal Bethe ansatz method \cite{Wan15}. Namely, an eigenvalue $\bar{\Lambda}(u)$ of the transfer matrix $t(u;\{\frac{1}{4},\,-\frac{1}{4}\})$ given by (\ref{trans-per-8}) can be expressed  in terms of a homogeneous T-Q relation
\bea
&& \bar{\Lambda}(u)=e^{i\phi}a(u)\frac{Q(u-\eta)}{Q(u)}+e^{-i\phi}d(u)\frac{Q(u+\eta)}{Q(u)}, \label{T-Q-1}\\[6pt]
&&a(u)=\frac{\s^{N}(u+\frac{1}{4}+\eta)\,\s^{N}(u-\frac{1}{4}+\eta)}{\s^{N}(\eta)\,\s^{N}(\eta)},
\quad d(u)=\frac{\s^{N}(u+\frac{1}{4})\,\s^{N}(u-\frac{1}{4})}{\s^{N}(\eta)\,\s^{N}(\eta)},\\[6pt]
&&Q(u)=\prod_{j=1}^{N}\frac{\s(u-\lambda_j)}{\s(\eta)},\label{T-Q-3}
\eea
where the $N+1$ parameters $\phi$ and $\{\lambda_j|j=1,\cdots,N\}$ have to satisfy the associated Bethe ansatz equations (BAEs)
\bea
&&e^{2i\phi}\,\frac{\s^{N}(\lambda_j+\frac{1}{4}+\eta)\,\s^{N}(\lambda_j-\frac{1}{4}+\eta)}{\s^{N}(\lambda_j+\frac{1}{4})\,\,\s^{N}(\lambda_j-\frac{1}{4})} =
\prod_{k\ne j}^N\frac{\s(\lambda_j-\lambda_k+\eta)}{\s(\lambda_j-\lambda_k-\eta)},\quad j=1,\cdots,N,\label{BAE-1}\\[6pt]
&&N\eta+2\sum_{j=1}^{N}\lambda_j=m_1+m_2\tau,\quad m_1,\,\,m_2\,\in Z.\label{BAE-2}
\eea
The whole set of solutions to the above BAEs can be decomposed into the classes labeled by  positive integers $\{k|k=0,1,\cdots,N-1\}$ according to the spectrum of $\tau(0)$
\bea
e^{2i\phi}\frac{Q(\frac{1}{4}+\eta)Q(-\frac{1}{4}-\eta)}{Q(\frac{1}{4})Q(-\frac{1}{4})}=e^{\frac{2i\pi k}{N}},\quad k=0,1,\cdots,N-1.\label{select-rule}
\eea
The detailed proof of (\ref{T-Q-1})-(\ref{select-rule}) will be  given in Appendix C.

Finally eigenvalues of the transfer matrix (\ref{trans-per}) of the closed elliptic $D^{(2)}_2$ spin chain can be obtained via the relation (\ref{eigenvaules}). Hence eigenvalue $E$ of the
Hamiltonian (\ref{ham})-(\ref{PBC}) can be expressed as
\bea
E&=&\lt.\Theta_0\frac{\partial}{\partial u}\ln \bar{\Lambda}(u+\frac{1}{4})\rt|_{u=0} +\lt.\Theta_0\frac{\partial}{\partial u}\ln \bar{\Lambda}(u-\frac{1}{4})\rt|_{u=0},\nonumber\\[6pt]
&=&-\Theta_0\sum_{j=1}^N\lt\{\zeta(\lambda_j+\eta+\frac{1}{4})+\zeta(\lambda_j+\eta-\frac{1}{4})-\zeta(\lambda_j+\frac{1}{4})-\zeta(\lambda_j-\frac{1}{4})\rt\}\nonumber\\[6pt]
&&\quad\quad+N\Theta_0\lt(\zeta(\eta+\frac{1}{2})+\zeta(\eta-\frac{1}{2})+2\,\zeta(\eta)\rt).\label{ham_sun}
\eea
where the constant $\Theta_0$ is given by (\ref{Int-R}), the function $\zeta(u)$ is defined in (\ref{Function}) and the parameters $\{\lambda_j|j=1,\cdots,N\}$ satisfy the BAEs (\ref{BAE-1})-(\ref{BAE-2}).

We have checked that the solution obtained by solving the BAEs (\ref{BAE-1})-(\ref{BAE-2}) indeed gives complete eigenvalues of the Hamiltonian of (\ref{ham})-(\ref{PBC}) by direct diagonalization numerically for small length $N$ of the lattice. For an example of $N=2$, the resulting Bethe roots and energies are listed in Table~\ref{tab:N2-bethe-energy}.

\begin{table}[htbp]
	\centering
	\caption{Numerical solutions of the BAEs
		(\ref{BAE-1})-(\ref{BAE-2}) for $N=2$, $\eta=0.19$,
		and $\tau=0.65i$. The energies $E_n$
		obtained from Eq.~(\ref{ham_sun}) agree with
		those from exact diagonalization.
		The index $n$ labels the distinct
		energy levels.}
	\label{tab:N2-bethe-energy}
	\begin{tabular}{cc|c|c|r|c}
		\hline\hline
		$\lambda_1$ & $\lambda_2$ & $\phi$ & $k$ & $E_n$ & $n$ \\
		\hline
		$0.15500000$ & $0.65500000$ & $0.00000000$ & $0$ & $-27.03795534$ & $1$ \\
		$0.25000000$ & $0.06000000$ & $0.00000000$ & $1$ & $-9.33068958$ & $2$ \\
		$0.15500000$ & $0.15500000+0.32500000i$ & $0.59690260$ & $1$ & $-9.33068958$ & $2$ \\
		$0.15500000$ & $0.65500000+0.32500000i$ & $0.59690260$ & $1$ & $-9.33068958$ & $2$ \\
		$0.75000000$ & $0.56000000$ & $0.00000000$ & $1$ & $-9.33068958$ & $2$ \\
		$0.65500000+0.32500000i$ & $0.65500000$ & $0.59690260$ & $1$ & $-9.33068958$ & $2$ \\
		$0.65500000$ & $0.15500000+0.32500000i$ & $0.59690260$ & $1$ & $-9.33068958$ & $2$ \\
		$0.65500000+0.32500000i$ & $0.15500000+0.32500000i$ & $1.19380521$ & $0$ & $8.37657618$ & $3$ \\
		$0.90500000+0.32500000i$ & $0.40500000+0.32500000i$ & $1.19380521$ & $0$ & $11.52494726$ & $4$ \\
		$0.40500000$ & $0.90500000+0.32500000i$ & $0.59690260$ & $0$ & $19.30095627$ & $5$ \\
		$0.40500000+0.32500000i$ & $0.40500000$ & $0.59690260$ & $0$ & $19.30095627$ & $5$ \\
		$0.40500000+0.53855916i$ & $0.40500000+0.11144084i$ & $1.19380521$ & $0$ & $19.30095627$ & $5$ \\
		$0.90500000$ & $0.40500000+0.32500000i$ & $0.59690260$ & $0$ & $19.30095627$ & $5$ \\
		$0.90500000+0.32500000i$ & $0.90500000$ & $0.59690260$ & $0$ & $19.30095627$ & $5$ \\
		$0.90500000+0.53855916i$ & $0.90500000+0.11144084i$ & $1.19380521$ & $0$ & $19.30095627$ & $5$ \\
		$0.40500000$ & $0.90500000$ & $0.00000000$ & $0$ & $27.07696527$ & $6$ \\
		\hline\hline
	\end{tabular}
	\par\smallskip
\end{table}


\section{Conclusions}
\label{Con}

An integrable elliptic generalization of the closed $D^{(2)}_2$ quantum spin chain is constructed by proposing the corresponding R-matrix $\mathbb{R}(u)$ given by (\ref{r-matrix})-(\ref{r-func}), which satisfies the QYBE (\ref{QYB}) and indeed ensures the integrability of the model described by the NN local interaction (\ref{ham})-(\ref{PBC}). Taking the trigonometric limit, we have shown that the corresponding R-matrix reduces to that
of the $D^{(2)}_2$ one (\ref{R-matrix-relation}), up to a similarity transformation. The very fact of the factorization for the R-matrix and
the corresponding transfer matrix allow us to give the exact solution of the model, which can be expressed in terms of
a homogeneous T-Q relation and the associated BAEs (\ref{T-Q-1})-(\ref{select-rule}).

\section*{Acknowledgments}

The author Wen-Li Yang would like to thank Xin Zhang for his useful discussions. The financial supports from National Program for Basic Research of
MOST (Grant No. 2021YFA1402104), National
Natural Science Foundation of China (Grant Nos. 12675004, 12675005, 12675006, 12547107, 12434006,
12305005 and 12247103),
Major Basic Research Program of Natural Science of Shaanxi
Province (Grant Nos. 2021JCW-19 and 2017ZDJC-32), Strategic Priority Research
Program of the Chinese Academy of Sciences, and Double First-Class
University Construction Project of Northwest University are
gratefully acknowledged.

\section*{Appendix A: Identities of the elliptic functions }
\setcounter{equation}{0}
\renewcommand{\theequation}{A.\arabic{equation}}

In this appendix, we list some identities of the elliptic functions which have been
used in the derivations of the paper. The elliptic
functions defined by (\ref{Function-a-b})-(\ref{Function}) satisfy the following identities:
\bea
\s(2u)&=&\frac{2\s(u)\s(u+\frac{1}{2})\s(u+\frac{\tau}{2})
\s(u-\frac{1}{2}-\frac{\tau}{2})}{\s(\frac{1}{2})\s(\frac{\tau}{2})
\s(-\frac{1}{2}-\frac{\tau}{2})},\label{Identity-f-1}\\[6pt]
\frac{\s(u)}{\s(\frac{\tau}{2})}&=&\frac{ \theta\lt[\begin{array}{l}
0\\\frac{1}{2}\end{array}\rt](u,2\tau)\,\,\theta\lt[\begin{array}{l}
\frac{1}{2}\\[2pt]\frac{1}{2}\end{array}\rt](u,2\tau)}
{\theta\lt[\begin{array}{l}
0\\\frac{1}{2}\end{array}\rt](\frac{\tau}{2},2\tau)\,\,
\theta\lt[\begin{array}{l}
\frac{1}{2}\\[2pt]\frac{1}{2}\end{array}\rt]
(\frac{\tau}{2},2\tau)},\label{Identity-f-2} \\[6pt]
\theta
\lt[\begin{array}{c}\frac{1}{2}\\[2pt]\frac{1}{2}\end{array}\rt]
(2u,2\tau)&=&\theta
\lt[\begin{array}{c}\frac{1}{2}\\[2pt]\frac{1}{2}\end{array}\rt]
(\tau,2\tau)\,\times\,\frac{\s(u)\s(u+\frac{1}{2})}
{\s(\frac{\tau}{2})\s(\frac{1}{2}+\frac{\tau}{2})},\label{Identity-f-3}\\[6pt]
\theta \lt[\begin{array}{c}0\\\frac{1}{2}\end{array}\rt]
(2u,2\tau)&=&\theta
\lt[\begin{array}{c}0\\\frac{1}{2}\end{array}\rt]
(0,2\tau)\,\times\,\frac{\s(u-\frac{\tau}{2})\s(u+\frac{1}{2}+\frac{\tau}{2})}
{\s(-\frac{\tau}{2})\s(\frac{1}{2}+\frac{\tau}{2})},\label{Identity-f-4}\\[6pt]
\sigma(u+\tau)&=&-e^{-2i\pi(u+\frac{\tau}{2})}\sigma(u),\quad
\sigma(u+1)=-\sigma(u).\label{Identity-f-5}
\eea
The
$\s$-function also satisfies the following
identity:\bea
&&\s(u+x)\s(u-x)\s(v+y)\s(v-y)-\s(u+y)\s(u-y)\s(v+x)\s(v-x)\no\\
&&~~~~~~=\s(u+v)\s(u-v)\s(x+y)\s(x-y),\label{Identity-fuction}\eea which
has been useful in deriving equations in this paper.

\section*{Appendix B: Limit to the $D^{(2)}_2$ R-matrix }
\setcounter{equation}{0}
\renewcommand{\theequation}{B.\arabic{equation}}

In this appendix, we consider the trigonometric limit
$\tau\rightarrow +i\infty$ of the R-matrix $\mathbb{R}(u)$ given by (\ref{r-matrix})-(\ref{r-func}) with the help of the product form (\ref{R-Factor}). For this purpose, let us first consider the trigonometric limit of the
eight-vertex R-matrix $R(u)$ given by (\ref{r-matrix-B})-(\ref{r-func-B}).  The definition of the
elliptic functions (\ref{Function-a-b})-(\ref{Function}) imply
\bea \s(u+\frac{\tau}{2})=e^{-i\pi(u+\frac{1}{2}+\frac{\tau}{4})}
\,\,\theta\lt[\begin{array}{c}0\\\frac{1}{2}\end{array}\rt](u,\tau),
\no\eea and the asymptotic behaviors
\bea
&&\lim_{\tau\rightarrow
+i\infty}\s(u)=-2e^{\frac{i\pi\tau}{4}}\sin \pi
u+\ldots,\no\\
&&\lim_{\tau\rightarrow
+i\infty}\theta\lt[\begin{array}{c}0\\\frac{1}{2}\end{array}\rt](u,\tau)=
1+\ldots.
\eea
The above asymptotic behaviors lead to
the well-known result:
\bea
\hspace{-1.2truecm}\lim_{\tau\rightarrow
+i\infty}R(u)=\frac{1}{\sin\,\pi\eta}\lt(\begin{array}{cccc}\sin\pi(u+\eta)
&&&\\&\sin\,\pi u&\sin\,\pi\eta&\\
&\sin\,\pi\eta&\sin\,\pi u&\\&&&\sin\pi(u+\eta)\end{array}\rt).\label{PR-matrix}
\eea
After re-parameterizing (\ref{Repar}), denoted the resulting limit R-matrix as $R^{(P)}(u)$, we have
\bea
R^{(P)}(u)=\frac{1}{\sinh\eta}\lt(\begin{array}{cccc}\sinh(\eta-\frac{u}{2})
&&&\\&\sinh\frac{u}{2}&\sinh\eta&\\
&\sinh\eta&\sin\frac{u}{2}&\\&&&\sinh(\eta-\frac{u}{2})\end{array}\rt).\label{PR-matrix-1}
\eea
Let us introduce the R-matrix $R^{(H)}(u)$
\bea
R^{(H)}_{12}(u)&=&\sinh\eta\,g_1(u)\,g_2(0)R^{(P)}_{12}(u)\,g^{-1}_1(u)\,g^{-1}_2(0)\nonumber\\[6pt]
&=&\lt(\begin{array}{cccc}\sinh(\eta-\frac{u}{2})
&&&\\&\sinh\frac{u}{2}&e^{-\frac{u}{2}}\sinh\eta&\\
&e^{\frac{u}{2}}\sinh\eta&\sin\frac{u}{2}&\\&&&\sinh(\eta-\frac{u}{2})\end{array}\rt),\label{PR-matrix-2}
\eea
where $g(u)={\rm diag}(e^{-\frac{u}{4}},e^{\frac{u}{4}})$. The product form \cite{Mar99, Fra12, Rob20, Nep21} of the $D^{(2)}_2$ R-matrix reads \cite{Nep21}
\bea
\mathbb{R}^{D^{(2)}_2}_{\hollow{1}\,\hollow{2}}(u)\equiv \mathbb{R}^{D^{(2)}_2}_{1\bar{1}\,2\bar{2}}(u)=2^4B_{1\bar{1}}B_{2\bar{2}}R^{(H)}_{1\,\bar{2}}(u+i\pi)\,R^{(H)}_{1\,2}(u)\,R^{(H)}_{\bar{1}\,\bar{2}}(u)\,
R^{(H)}_{\bar{1}\,2}(u-i\pi)B_{1\bar{1}}B_{2\bar{2}},
\eea
which directly leads to the limiting relation (\ref{R-matrix-relation}). Here, the similarity transformation $B$ is
\bea
B= \lt(\begin{array}{llll}1&&&\\[6pt]
&\frac{\cosh\frac{\eta}{2}}{\sqrt{\cosh\eta}}&-\frac{\sinh\frac{\eta}{2}}{\sqrt{\cosh\eta}}&\\[6pt]
&-\frac{\sinh\frac{\eta}{2}}{\sqrt{\cosh\eta}}&-\frac{\cosh\frac{\eta}{2}}{\sqrt{\cosh\eta}}&\\[6pt]
&&&1
\end{array}\rt),\quad B^2={\rm id}.
\eea

\section*{Appendix C: T-Q relation of the closed XYZ  chain}
\setcounter{equation}{0}
\renewcommand{\theequation}{C.\arabic{equation}}

In this appendix, following the method developed in \cite{Wan15}, we construct a T-Q relation of eigenvalue $\bar{\Lambda}(u)$ of the transfer
matrix $t(u)$ of the inhomogeneous closed XYZ spin-$\frac{1}{2}$ chain. Let us introduce the transfer matrix $t(u)$ of inhomogeneous closed XYZ chain
with the length $L$ as
\bea
t(u)=tr_{0}\lt\{R_{0L}(u-\theta_L)R_{0L-1}(u-\theta_{L-1})\cdots R_{01}(u-\theta_1)\rt\},\nonumber
\eea
where $\{\theta_j|j=1,\cdots,L\}$ are the inhomogeneous  parameters, and the R-matrix $R(u)$ is the eight-vertex R-matrix (\ref{r-matrix-B}). An eigenvalue $\bar{\Lambda}(u)$ of the transfer matrix $t(u)$ generally can be given in terms of
an inhomogeneous T-Q relation \cite{Cao1, Cao14}
\bea
\bar{\Lambda}(u)&=&e^{2i\pi l_1u+i\phi}a_L(u)\frac{Q_1(u-\eta)}{Q_2(u)}+e^{-2i\pi l_1(u+\eta)-i\phi}d_L(u)\frac{Q_2(u+\eta)}{Q_1(u)}\nonumber\\[6pt]
&&\quad\quad+c\frac{\s^m(u+\frac{\eta}{2})\,a_{L}(u)\,d_{L}(u))}{\s^m(\eta)\,Q_1(u)\,Q_2(u)},\label{T-Q-1-B}\\
a_L(u)&=&\prod_{l=1}^L\frac{\s(u-\theta_l+\eta)}{\s(\eta)},\quad
d_L(u)=\prod_{l=1}^L\frac{\s(u-\theta_l)}{\s(\eta)},\label{T-Q-2-B}\\[6pt]
Q_1(u)&=&\prod_{j=1}^M\frac{\s(u-\mu_j)}{\s(\eta)},\quad  Q_2(u)=\prod_{j=1}^M\frac{\s(u-\nu_j)}{\s(\eta)},
\eea
where $l_1$ is an integer, $\phi$ and $c$ are two complex parameters, and $M$ and $m$ are two non-negative integers with a constraint: $L+m=2M$. The $2M+3$ parameters have to satisfy
the associated BAEs
\bea
&&\lt(\frac{L}{2}-M\rt)\eta-\sum_{j=1}^M(\mu_j-\nu_j)=l_1\tau+m_1,\quad m_1\in Z,\label{BAE-1-B} \\[6pt]
&&M\eta-\sum_{l=1}^L\theta_l+\sum_{j=1}^M(\mu_j+\nu_j)=m_2,\quad m_2\,\in Z,\label{BAE-2-B}\\[6pt]
&&c\,e^{2i\pi l_1(\mu_j+\eta)+i\phi}\frac{\s^m(\mu_j+\frac{\eta}{2})}{\s^m(\eta)}\,a_L(\mu_j)=-Q_2(\mu_j)Q_2(\mu_j+\eta),\quad j=1,\cdots,M,\label{BAE-3-B}\\[6pt]
&&c\,e^{-2i\pi l_1\nu_j-i\phi}\frac{\s^m(\nu_j+\frac{\eta}{2})}{\s^m(\eta)}\,d_L(\nu_j)=-Q_1(\mu_j)Q_1(\nu_j-\eta),\quad j=1,\cdots,M,\label{BAE-4-B}\\[6pt]
&&\prod_{l=1}^L\bar{\Lambda}(\theta_l)=\prod_{l=1}^La_L(\theta_l).\label{BAE-5-B}
\eea
For generic $\eta$, $\tau$ and an even $L$, the solutions of (\ref{BAE-1-B})-(\ref{BAE-5-B}) with a vanishing $c$ (i.e. $c=0$) may  require
\bea
l_1=0,\quad L=2M,\quad \{\mu_j\}=\{\nu_j\}\stackrel{{\rm denote}}{=} \{\lambda_j+m_j\tau\}\quad {\rm with}\quad m_j\in Z.\label{Even-L}
\eea
This gives rise to a homogeneous T-Q relation
\bea
&&\bar{\Lambda}(u)= e^{i\phi}a_L(u)\frac{Q(u-\eta)}{Q(u)}+e^{-i\phi}d_L(u)\frac{Q(u+\eta)}{Q(u)},\label{T-Q-3-B}\\
&&Q(u)=\prod_{j=1}^{M}\frac{\s(u-\lambda_j)}{\s(\eta)}\equiv \prod_{j=1}^{\frac{L}{2}}\frac{\s(u-\lambda_j)}{\s(\eta)}.\label{T-Q-4-B}
\eea
Now the associated  BAEs (\ref{BAE-1-B})-(\ref{BAE-4-B}) reduce to \footnote{The condition (\ref{Even-L}) gives rise to that the BAE (\ref{BAE-2-B}) becomes (\ref{BAE-6-B}). It is easy to check that under the condition (\ref{Even-L}), the BAE (\ref{BAE-8-B}) is actually a simple consequence of the resulting   BAEs (\ref{BAE-7-B}), but now becomes the classification of the set of solutions to the BAEs (\ref{BAE-1})-(\ref{BAE-2}) such as (\ref{select-rule}) instead.}
\bea
&&M\eta-\sum_{l=1}^L\theta_l+2\sum_{j=1}^M\lambda_j=m_2+m_3\tau,\quad m_2,\,m_3\in Z,\quad M=\frac{L}{2},\label{BAE-6-B}\\[6pt]
&&e^{2i\phi}\frac{a_L(\lambda_j)}{d_{L}(\lambda_j)}=-\frac{Q(\lambda_j+\eta)}{Q(\lambda_j-\eta)},\quad j=1,\cdots,M=\frac{L}{2},\label{BAE-7-B}\\[6pt]
&&\prod_{l=1}^L\bar{\Lambda}(\theta_l)=\prod_{l=1}^La_L(\theta_l).\label{BAE-8-B}
\eea
With the special choices of the length of the sites $L=2N$, and the inhomogeneous parameters $\theta_{2l-1}=\frac{1}{4},\,\theta_{2l}=-\frac{1}{4},\,l=1,\cdots N$, the resulting T-Q relation (\ref{T-Q-3-B})-(\ref{T-Q-4-B}) and the associated BAEs  (\ref{BAE-6-B})-(\ref{BAE-8-B})
lead to (\ref{T-Q-1})-(\ref{BAE-2}).

\end{document}